\documentclass[intlimits,twoside,a4paper]{article}

\usepackage{amsmath,amssymb}
\usepackage{graphicx}

\usepackage[T2A]{fontenc}
\usepackage[cp1251]{inputenc}

\usepackage[eqsecnum]{cmpj3}

\issue{2017}{20}{3}{33003}
\doinumber{10.5488/CMP.20.33003}
\title[Destruction processes in self-irradiated materials]%
{Stochastic simulation of destruction processes in self-irradiated materials\thanks{We 
dedicate this article to the memory of late Dr. Jean-Pierre Badiali who
inspired our research and made a significant contribution to this work.}}
\author[T.~Patsahan \textsl{et al.}]{T.~Patsahan\refaddr{icmp},
        A.~Taleb\refaddr{psl,upmc}, J.~Stafiej\refaddr{cswu}, M.~Holovko\refaddr{icmp},
        \framebox{J.P.~Badiali\refaddr{icmp,upmc}}}
\addresses{
\addr{icmp} Institute for Condensed Matter Physics of the National Academy of Sciences of Ukraine,\\
1~Svientsitskii St., 79011 Lviv, Ukraine
\addr{psl} PSL Research University, Chimie ParisTech --- CNRS, Institut de Recherche de Chimie Paris, Paris 75005, France
\addr{upmc} Universit\'{e} Pierre et Marie Curie, Paris, 75231, France
\addr{cswu} Cardinal Stefan Wyszy\'{n}ski University, Department of Mathematics and Natural Sciences, Warsaw, Poland
}

\date{Received June 19, 2017, in final form August 6, 2017}

\begin{document}

\maketitle

\begin{abstract}
Self-irradiation damages resulting from fission processes are common phenomena observed in nuclear fuel containing (NFC) materials. Numerous $\alpha$-decays lead to local structure transformations in NFC materials.
The damages appearing due to the impacts of heavy nuclear recoils in the subsurface layer can cause detachments of material particles. Such a behaviour is similar to sputtering processes observed during a bombardment of the material surface by a flux of energetic particles. However, in the NFC material, the impacts are initiated from the bulk.
In this work we propose a two-dimensional mesoscopic model to perform a stochastic simulation of the destruction processes occurring in a subsurface region of NFC material.
We describe the erosion of the material surface, the evolution of its roughness and predict the detachment of the material particles. Size distributions of the emitted particles are obtained in this study. The simulation results of the model are in a qualitative agreement with the size histogram of particles produced from the material containing lava-like fuel 
formed during the Chernobyl nuclear power plant disaster.
\pacs 02.50.-r, 23.70.+j, 05.50.+q, 46.50.+a, 46.65.+g
\end{abstract}

\section{Introduction}
A study of nuclear fuel containing (NFC) materials and problems related to
their storage are of a special interest due to the great
importance with respect to the safety of environment.
The nuclear waste materials encapsulating actinides (e.g., U, Pu) undergo
self-irradiation during long periods of time. Therefore,
their chemical and physical properties change crucially.
Permanent $\alpha$-decay events occurring in a material lead to transformations of material structure around
the points where these events appear, producing the so-called thermal spikes~\cite{matzke,weber,weber2,ewing}.
Such local structural changes damage a material and can lead to its destruction if the relaxation time related to healing processes is too slow.
In homogeneous NFC materials where nuclear fuel is distributed randomly, the damages also appear randomly in the material bulk producing local defects \cite{weber2}. On the other hand, if these defects occur deep in the bulk they can be healed with time~\cite{weber2,ewing}, while the defects produced near the material surface can erode this surface which leads to a detachment of the particles formed by material pieces.
As a consequence, a highly radioactive surface may spontaneously disseminate and disperse a solid state phase into the environment. In connection with these phenomena, long term predictions are required.
However, predictions cannot result from simplified extrapolations of
the experimental data measured in limited time scales, which are much shorter than the storage period of the nuclear waste. Therefore, theoretical description appears to be an unavoidable complement to experimental data, and computer simulations become helpful in this case.

There is a large literature describing the damage processes inside nuclear waste materials
\cite{evron,fayek,trachenko,trachenko2,trachenko3,bishop,brutzel,brutzel2,martin,martin2,tian,delaye,dewan,kieu,devanathan,heinisch,chung,baclet,dremov,robinson2,geisler}. Self-destruction mechanisms of pure nuclear fuel are also studied. Usually, they try to predict the conditions under which a material bulk becomes unstable (fragile). In most theoretical studies, an atomistic level of modelling is used.
For instance, in \cite{trachenko2} a method of molecular dynamics is used to simulate the damage processes in zirconolite structures containing radioactive Pu. Unfortunately, such studies cannot be performed on large time scales as
well as one cannot make any predictions for the system where many nuclear decays occur, since the microscopic level description imposes a restriction on a space scale as well. Therefore, it is necessary to
develop a model allowing us to consider the system at a mesoscopic level.

In the present study, we propose a simple mesoscopic model of surface destruction processes in NFC materials,
which is implemented on two-dimensional lattice with the use of a stochastic algorithm similar
to the cellular automata concept~\cite{nous2,nous3}.
It is assumed that the detachment of nanoparticles
at the NFC material surface results from the formation of damaged regions and microfracturing in the vicinity of
the locations where nuclear decays take place.
It is also assumed that the heavy nuclear recoils due to $\alpha$-decays
are responsible for the formation of the damaged regions inside the NFC material.
The technique used in the study allows us to run simulations at a mesoscopic level and to
describe the surface evolution in large time scales. Thus, the characteristics such as chemical roughness, the mean-square front width and the yield of a material can be easily obtained from simulations.
Our main goal is to get histograms of the particles formed during the surface destruction of NFC material.
We compare the obtained histograms with the experimental results obtained
for the NFC material formed in the well-known heavy nuclear accident at the Chernobyl nuclear power plant~\cite{baryakhtar}.

This paper is organized as follows.
In section~\ref{sec2}, we discuss the physical aspects and review some literature related
to the damage and destruction processes in NFC materials.
A stochastic model developed in our study is described in section~\ref{sec3}.
In section~\ref{sec4}, we present a discussion of the obtained results.
Finally, we conclude the paper in the last section.

\section{Physical aspects}
\label{sec2}
To date, many studies can be found in literature which are devoted to
the study of the damage and destruction processes of the solid material surface due to
\textit{external} factors such as a high energy particle
bombardment, while there is a lack of theoretical studies of the same processes
for the NFC materials where the \textit{internal} factors cause a surface destruction.
However, one can notice similarities between the processes occurring
near the surface in the self-irradiated material and the
sputtering processes due to a surface bombardment with a flux of energetic particles. In the case of
collisions between the beam and the target, one can observe a formation of cascades of atom displacements,
and as a consequence, the creation of new structures and defects inside the material, which lead to a surface
reconstruction and particles emission.
For energetic incident beams, the sputtered particles can be large clusters of atoms.
These clusters are formed in the near-surface region due to a multiple breaking of bonds~\cite{wucher,betz}.
The magnitude of these effects specifically depends on the nature of
the particles forming the beam and on their energy as well as on the
nature of a target material. When a target and a beam contain the same
chemical species, this process is called a self-sputtering.
The self-sputtering is also widely studied for materials bombarded with
projectiles of different energies \cite{robinson,voskoboinikov,salonen} and sizes \cite{li,lindenblatt}.

In the case of sputtering phenomena as well as in the case of
self-irradiated materials, we deal with energy dissipation, which leads to the
formation of atomic displacement cascades \cite{ewing, betz}. The
main difference is that in the first case, the cascades are
initiated at the surface, and the atoms or atomic clusters escape from
the surface immediately. In the second case, the cascades
start in the bulk.
Due to $\alpha$-decays occurring in NFC material and a heavy nuclear recoil appears. As a consequence, 
an energy in the range of $40{-}100$~keV disperses causing
a local structure transformation. As a result of elastic collisions
between atoms and heavy nuclei, the cascades of atomic displacements
around $\alpha$-decay locations must occur\cite{day,weber,ewing,trachenko,palenik}.
The formation of atomic displacement cascades produced by an elastic collision
with some energetic particles was studied by molecular dynamics
\cite{trachenko,trachenko2,kubota1} and also observed from experiment \cite{kubota2}.
The sizes of atomic collision cascades depending on the value of the particle energy estimated in these investigations
are in the range of 25--30~nm. The same result was obtained by Baryakhtar  et al. \cite{baryakhtar} for
lava-like fuel containing materials (LFCM) formed in the
Chernobyl heavy nuclear accident~\cite{pazukhin,avzhidkov}. A microscopic
structure of a damaged region is disordered. At the same time, it is known that such
disordered regions cause local tensions, which are responsible for the defects and microfractures formation
in NFC materials \cite{weber2,ewing,evron,kieu} and they can crucially reflect on the material
durability \cite{ewing,day}. However,  the fractures produced far from
the material surface may not play an essential role in the material
destruction due to two factors: i) because they are confined in
the stable material parts; ii) they can be healed in a quite short
time due to relaxation processes \cite{wronkiewicz}. On the other
hand, when fractures appear close to the surface, their boundaries can
reach a surface and detach parts of the material. A size of
a detached particle depends on geometrical sizes of the fractures which are
defined by mechanical properties of the material considered as
well as by the collision energy.

\section{Modelling}
\label{sec3}
\subsection{General concept}
We propose a simple model sufficient to describe the cluster formation in terms of two
parameters model. First we assume that we have a lattice
with the sites from which nuclear events can be initiated.
When one of the events occurs at some given site, the damage randomly expands producing defects
in the vicinity of this site. For instance, a formation of a cascade of atom displacements (damaged region)
takes place.
In this case, one of the model parameters should relate to an area of a contact surface between
the damaged region and the rest of the material forming a fracture.
Another parameter corresponds to the frequency of nuclear events per volume of the NFC material,
which directly depends on the concentration of nuclear fuel.
A fracture can lead to the detachment of a piece of the material (particle detachment)
when both two conditions are valid:
i) boundaries of the fracture totally belong to the surface;
ii) the fracture is open toward the surface, e.g., it should have a hemisphere-like or a cap-like shape
with a basis directed toward the surface.
The sequential particle detachments lead to the formation of a specific front surface which is characterized
by its roughness.

To facilitate our simulations, we consider {a two-dimensional model of the system
presented as its projection}. Therefore, the NFC material is {described} by its profile in $XY$ coordinates,
where the surface destruction process proceeds along $Y$-axis. In this case, the area of a fracture is reduced to the length of its projection,
and for a particle detachment it should be of an arc-like shape. Moreover, the front surface formed during the destruction
process in two dimensions reduces to a line, and hereafter we refer to it as a front line (FL).

\subsection{Stochastic model}
\begin{figure}[!t]
\begin{minipage}{0.495\textwidth}
\begin{center}
\includegraphics[width=0.7\textwidth]{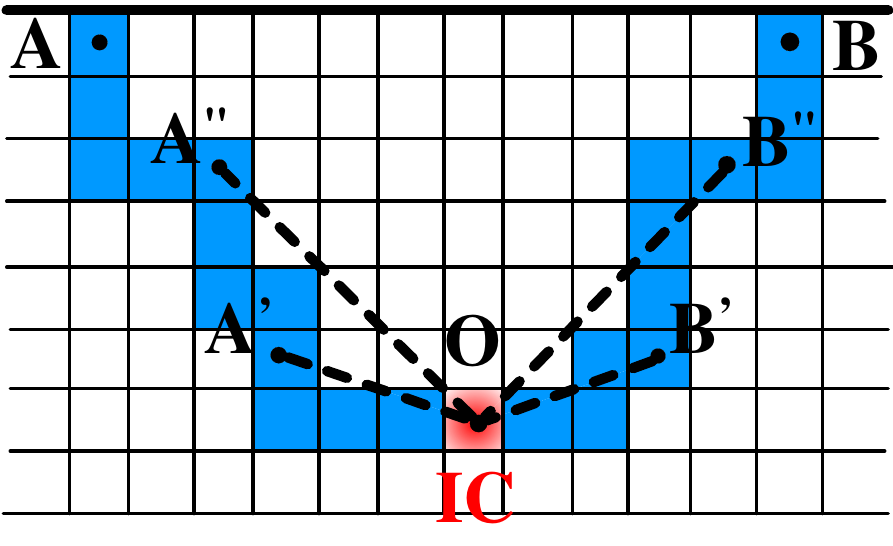}
\end{center}
\end{minipage}
\begin{minipage}{0.495\textwidth}
\begin{center}
\includegraphics[width=0.7\textwidth]{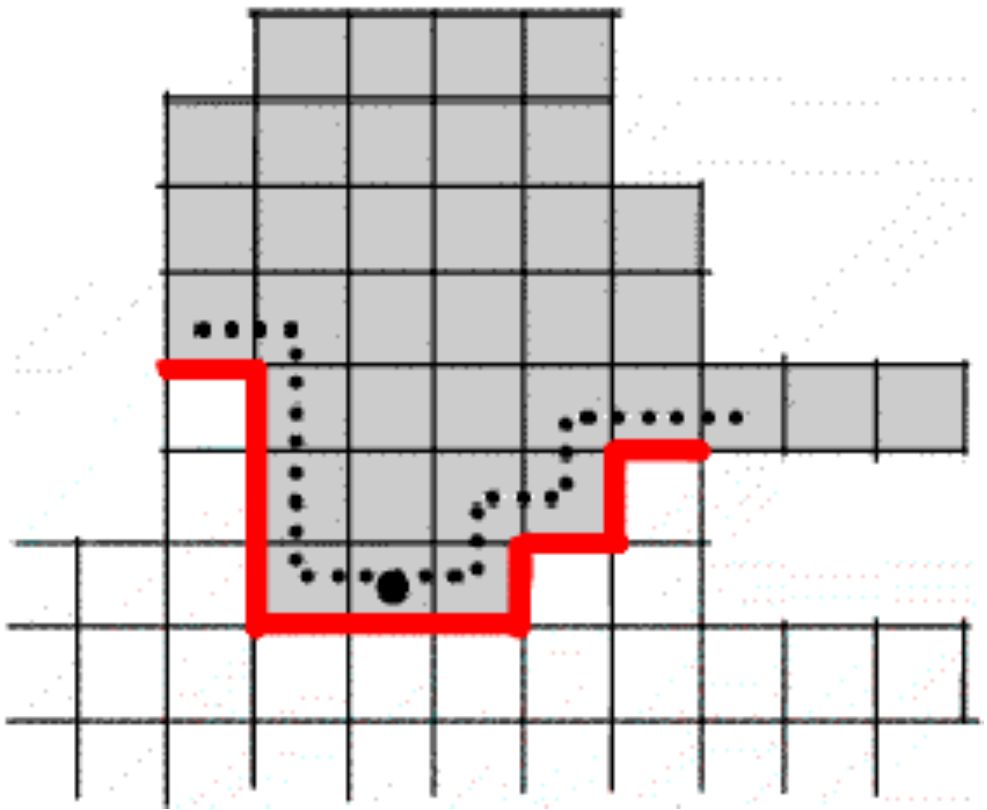}
\end{center}
\end{minipage}
\begin{minipage}{0.495\textwidth}
\vspace{-3.5mm}
\caption{\label{fig:scheme} (Color online) Example of fracture propagation.
Filled blue cells denote random walk paths to build a fracture.
The cental red cell (site ``O'') corresponds to the impact center (IC).}
\end{minipage}
\begin{minipage}{0.495\textwidth}
\caption{\label{fig:example} %
(Color online) Cluster detachment for $N_{\text{srw}}=6$. Filled grey cells belong to the
particle to be detached. The dotted lines denote random walk paths
initiated from the IC site (big point) and the thick red line denotes the
corresponding fracture.}
\end{minipage}
\end{figure}

A two-dimensional model is presented as a square
lattice with 4 or von Neumann connectivity \cite{neumann}. The
simulation box is of the size of $N_x\times N_y$ and the
periodical boundary conditions are applied along $X$-axis. The
sites of the lattice are of two types corresponding to two kinds
of species. The first kind of sites represents the
regions of pristine solid material and the second type represents the
regions containing nuclear fuel inclusions, which correspond to impact centers (ICs)
where the damage processes can start from.
We reduce the number of characteristic lengths of the system to only one ---
the lattice constant, $a$, and it is subsequently our unit of
length. The IC sites are located randomly on the lattice. We
denote their concentration or number density as $f_{\text d}$.
{Each of the IC sites} is chosen randomly to build paths of a fracture
initiated from this IC (OA and OB in figure~\ref{fig:scheme}).
Within the model proposed, the fracture is built by two different paths, which evolve
in opposite directions. Such a fracture formation can potentially lead to a detachment of a piece of material surface.
In this case, two possibilities to start the fracture creation are considered: OA is directed
to the left from the chosen IC site while OB goes to the right,
or OA goes up while OB goes down. The paths of fracture propagation
is modelled as a random walk. Due to the assumption of a limited
length of the fracture that can be formed in the material, the number
of steps of a random walk cannot exceed the specified parameter
$N_{\text{srw}}$, which is equal to a half the maximum length of fractures.
Another restriction introduced in the model requires that the
distances between the ends of the fracture and its initial point
should increase during its propagation as shown in figure~\ref{fig:scheme}:
$\text{OA}^{\prime\prime}>\text{OA}^{\prime}$ and
$\text{OB}^{\prime\prime}>\text{OB}^{\prime}$. This condition prevents
the self-crossing of the fracture. In this way, the ends of a fracture propagate away one from
another forming an arc-like fracture. It is quite reasonable
to give some preferable direction for the fracture propagation
dependent on a deformation geometry around the IC considered
(especially when the maximum length of a fracture is rather small). If
both ends of the fracture reach the surface, then the
constructed fracture leads to a detachment of the material
cluster~(figure~\ref{fig:example}). Otherwise, the event is discarded,
i.e., when the fracture paths stop somewhere in the bulk. In this case, we forget about it supposing that the
relaxation processes are fast enough for the fracture to be
healed~\cite{wronkiewicz}. To obtain the results with a sufficient
accuracy for the surface roughness for each pair of the parameters
$N_{\text{srw}}$ and $f_{\text d}$, we perform an average over $30$ independent
simulation runs. Each simulation run takes $200000$ steps.
{One simulation step corresponds to one particle detachment.}
The size of a lattice taken of $N_x=1000$ and $N_y=1000$ is found to be
sufficient to minimize the boundary condition effects {and to consider rather large FL widths}.
The events for the size histogram are collected in a stationary regime of the material
destruction after $50000$ simulation steps. {To identify the clusters of sites formed during particle detachments
the Hoshen-Kopelman algorithm is used \cite{Hoshen}}.

\section{Results and discussion}
\label{sec4}
To describe the properties of the front line (FL) we introduce standard functions, which
characterize the growth processes \cite{barabasi}.
First we consider the roughness of the FL as a function of the number of simulation time steps $N_{\text{steps}}$ for different values of the parameters $f_{\text d}$ and $N_{\text{srw}}$.
The roughness is given by the factor defined as a ratio of the number of sites $N_{\text s}$ in the FL to the initial number $N_{x}$ of sites in the smooth surface $N_x=1000$. After $N_{\text{step}}$, the roughness factor is given by
\begin{equation}
r(N_{\text{steps}}) = \frac{N_{\text s} (N_{\text{steps}} )}{N_x }\,. %
\label{rough}
\end{equation}

\begin{figure}[!t]
\begin{center}
\includegraphics[clip,width=0.495\textwidth,angle=0]{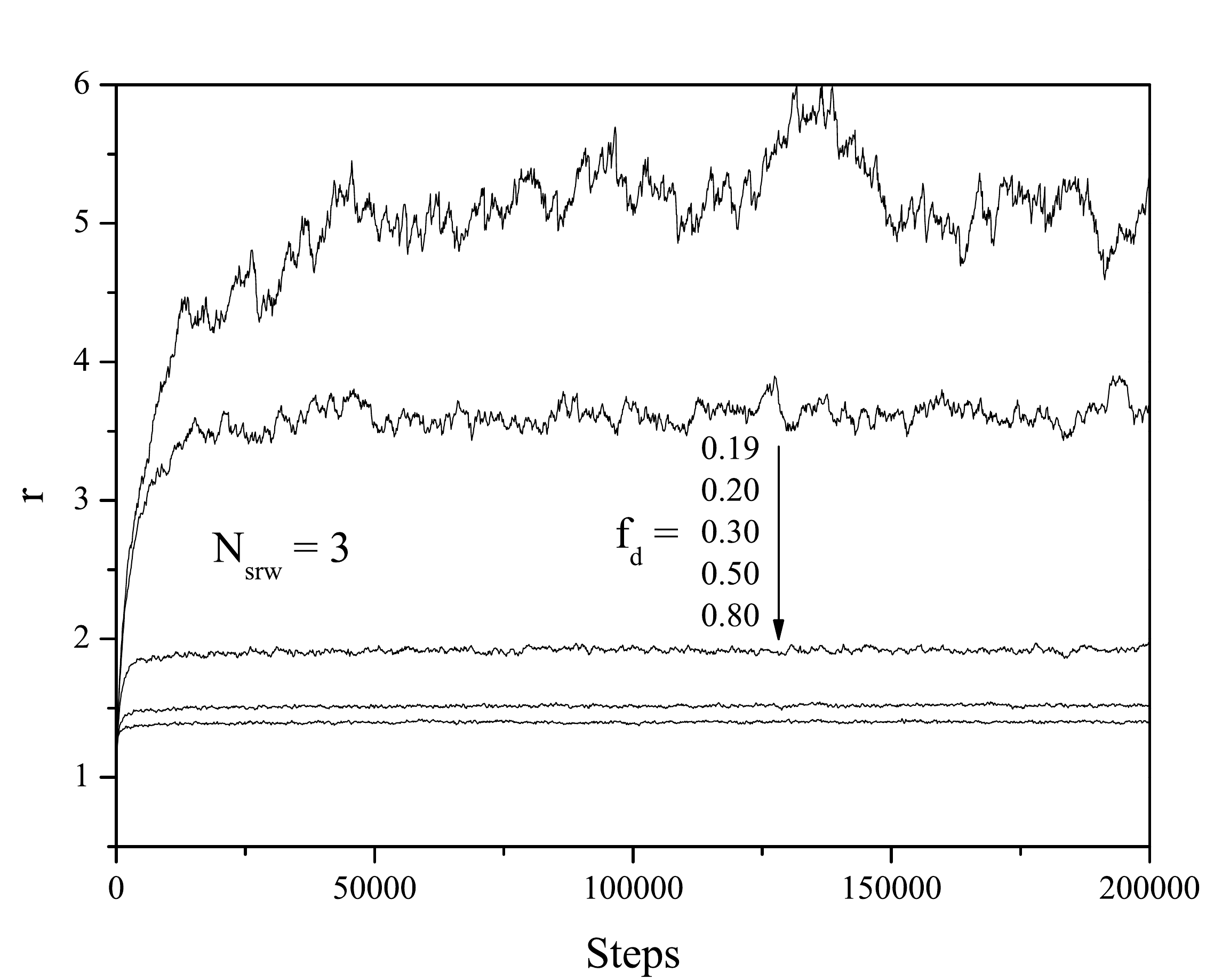}
\includegraphics[clip,width=0.495\textwidth,angle=0]{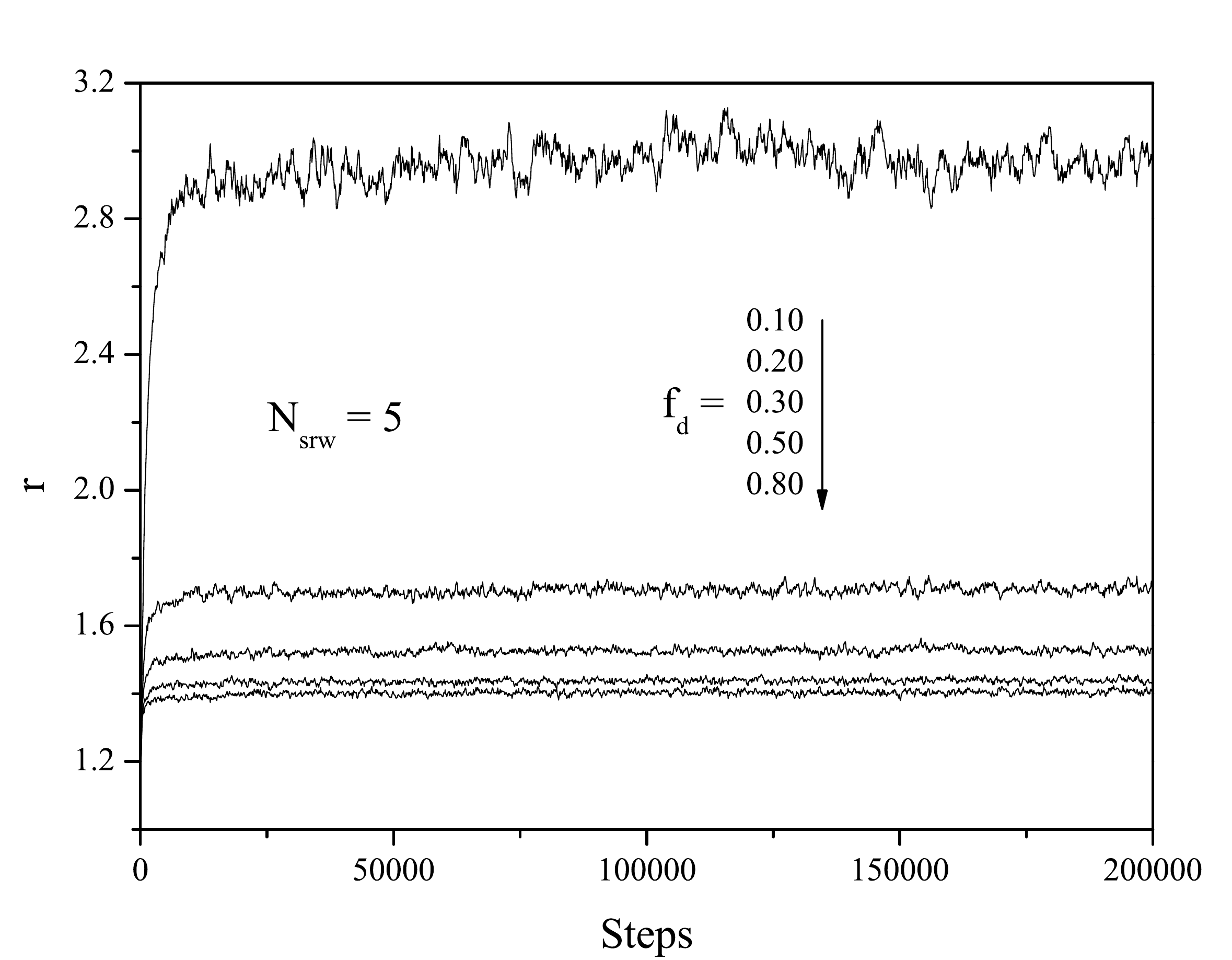} 
\caption{\label{fig:roughness} %
Variation of the ratio $r$ versus a number of simulation steps
obtained for $N_{\text{srw}}=3$ and $N_{\text{srw}}=5$, and for the different density $f_{\text d}$.}
\end{center}
\end{figure}
\begin{figure}[!t]
\vspace{-5mm}
\begin{center}
\includegraphics[clip,width=0.55\textwidth,angle=0]{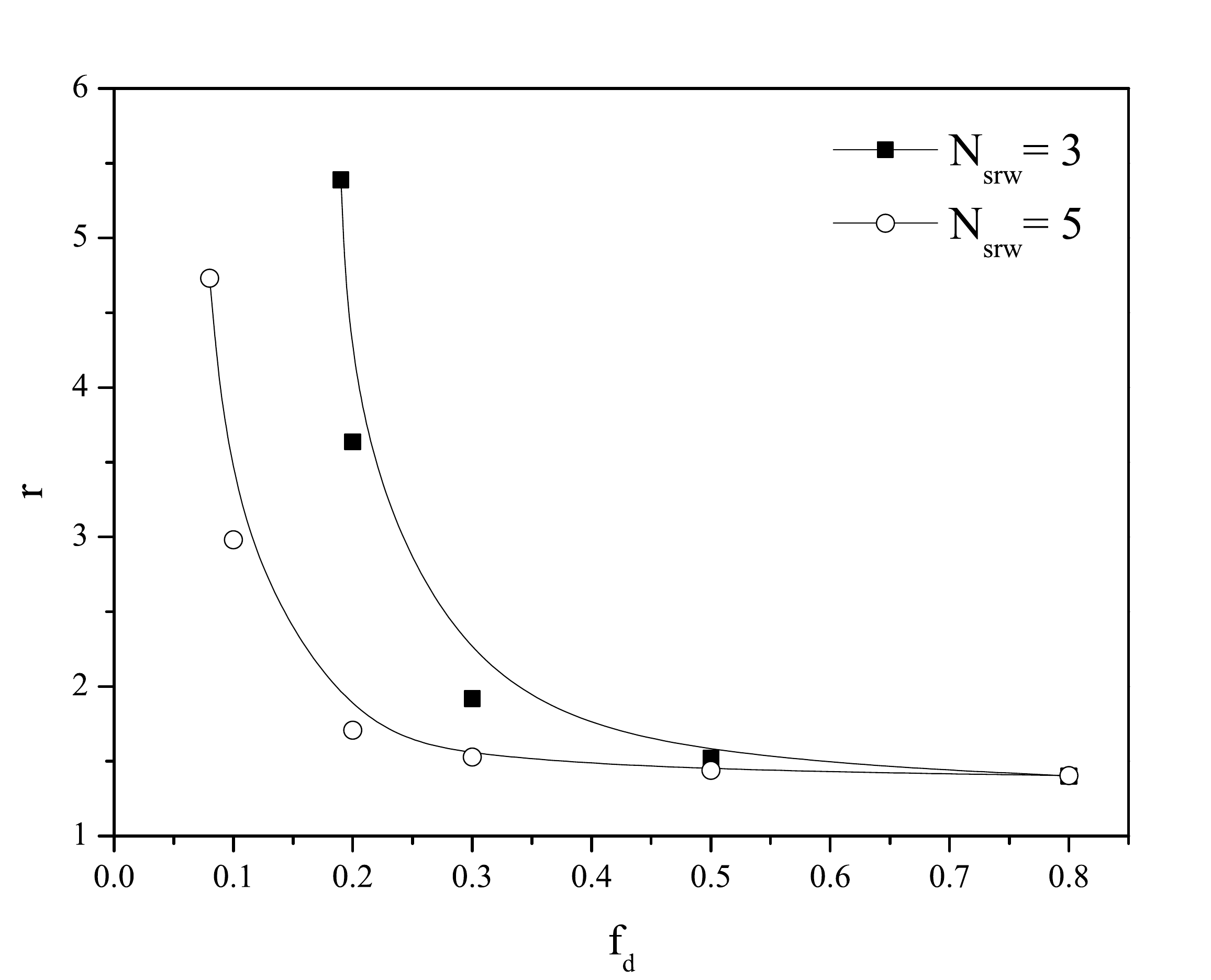} 
\caption{\label{fig:roughness2} The dependence of
roughness $r$ on the $f_{\text d}$ for $N_{\text{srw}}=3$ (square symbols)
and $N_{\text{srw}}=5$ (circle symbols).}
\end{center}
\vspace{-3mm}
\end{figure}

We may notice that {according to our algorithm}, the $N_{\text s}$ sites are only formed by the host material
(free of any nuclear fuel inclusions). Thus, there is no IC site on the surface, by construction.
For the initial FL there is no roughness, i.e., $r(0)=1$.
After some simulation steps, we observe an increase of the roughness factor, due
to the particle detachment process (figure~\ref{fig:roughness}).
After $50000$ simulation steps we get a stationary regime characterized
by a constant value $r$ for all the parameters considered in our study.
In figure~\ref{fig:roughness2} we plot $r$ as a function of $f_{\text d}$ and for two values of $N_{\text{srw}}$.
We also observe that $r$ increases when the density of ICs decreases and about $f_{\text d}=0.80$ 
the roughness becomes almost independent of $N_{\text{srw}}$. We expect that this weakening related to
the average distance between ICs becomes less than the mean length of a fracture.
This can be also explained by the fact that the mean size of detached particles increases monotonously when we decrease $f_{\text d}$ (table~\ref{tab:size}). It is also worth noting that there exists a critical value $f_{\text c}$ such that for $f_{\text d}<f_{\text c}$ the destruction process stops.
{This is related to the distances between ICs, which are too large in order to provide continuous detachments of particles,
and at some step one quickly faces a situation when $N_{\text{srw}}$ is not sufficient to build a path from any IC in the system
to the surface.}
For $N_{\text{srw}}=3$ we obtain $f_{\text c}$ about $0.19$, and for $N_{\text{srw}}=5$, the critical density $f_{\text c}$ is found at lower values of $f_{\text d}$ which are near $0.08$.
{One can observe in figure~\ref{fig:roughness2} that in the vicinity of $f_{\text c}$, the roughness $r$ increases drastically, and
since the critical density $f_{\text c}$ is higher for smaller $N_{\text{srw}}$, the roughness increases faster with a decrease of $f_{\text d}$.
Therefore, we always obtain larger values of the roughness for smaller $N_{\text{srw}}$. This looks surprising because one would expect a larger
roughness in the case corresponding to a larger particles detachment. This is certainly true if a single particle detachment
at a flat surface is considered. However, we have a series of detachments leading to a formation of a complex rough surface.
It is really difficult to predict for which exactly $N_{\text{srw}}$, the roughness is higher without doing simulation.
Moreover, it should be noted that $N_{\text{srw}}$ is just the maximum length of a fracture, and it means that actual lengths of fractures
causing particle detachments can be smaller than the given $N_{\text{srw}}$. Therefore, one can expect a distribution of lengths of such fractures. At the same time, the roughness results from the averaging along FL and the real contribution of fractures of different sizes to the total surface formation is unclear.
On the other hand, the behaviour of the roughness at different $N_{\text{srw}}$ can be explained by that for a larger $N_{\text{srw}}$ where the destruction process proceeds more intensively than for smaller $N_{\text{srw}}$, and this leads to a better smoothing of FL.}

\begin{table}[!t]
\begin{center}
\caption{\label{tab:size} The mean sizes of particles for the given parameters.}
\vspace{2ex}
\begin{tabular}
{|c|c|c|} \hline\hline $f_{\text d}$&~$N_{\text{srw}}= 3$~&
~$N_{\text{srw}}= 5$~ \\
\hline\hline ~0.10~& --&
14.78 \\
\hline ~0.20~& 10.06&
\textbf{13.58} \\
\hline ~0.30~& \textbf{9.72}&
\textbf{12.49} \\
\hline ~0.50~& \textbf{8.60}&
\textbf{11.70} \\
\hline ~0.80~& \textbf{8.05}&
\textbf{11.32} \\
\hline\hline
\end{tabular}
\end{center}
\end{table}

\begin{figure}[!b]
\begin{center}
\includegraphics[clip,width=0.495\textwidth,angle=0]{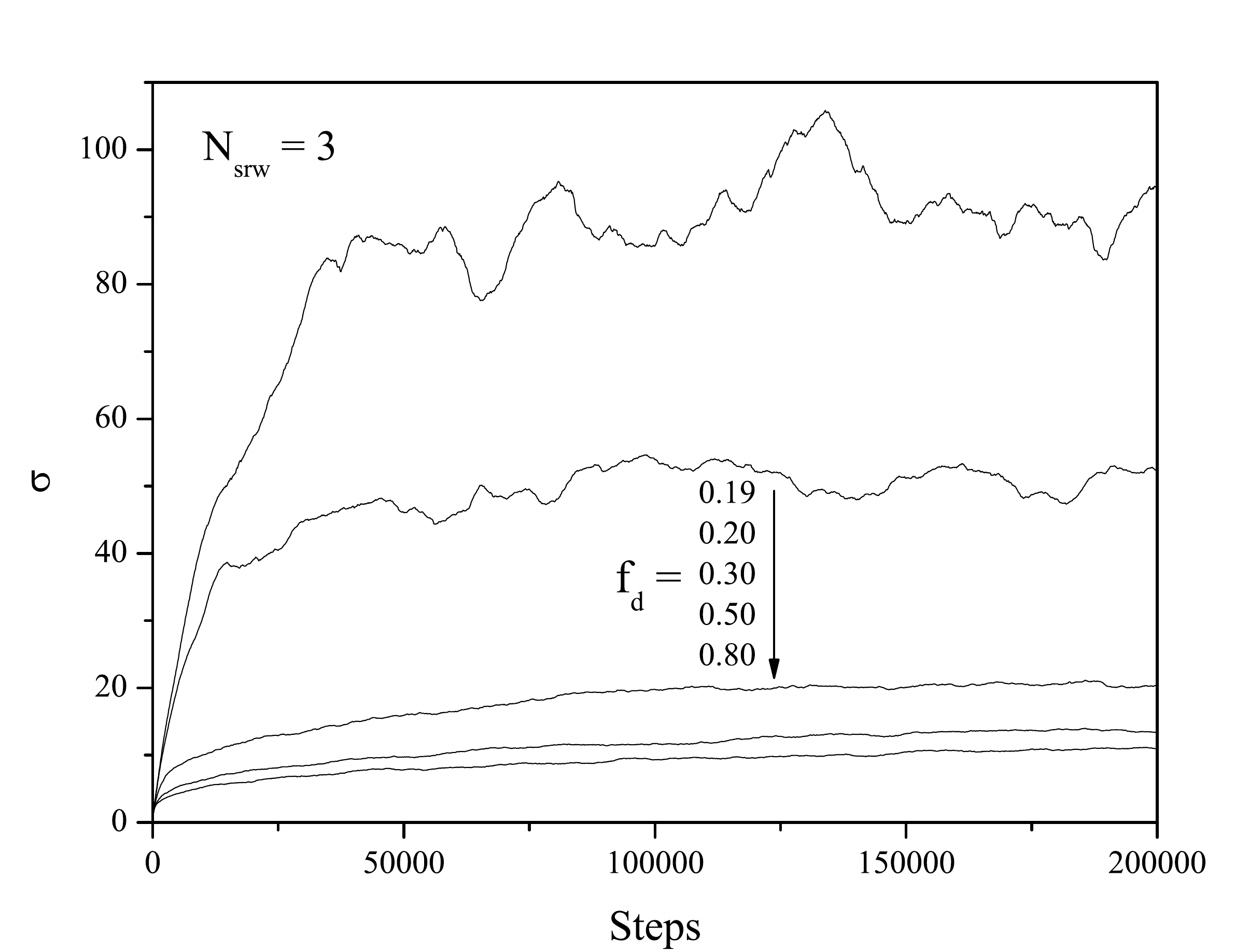}
\includegraphics[clip,width=0.495\textwidth,angle=0]{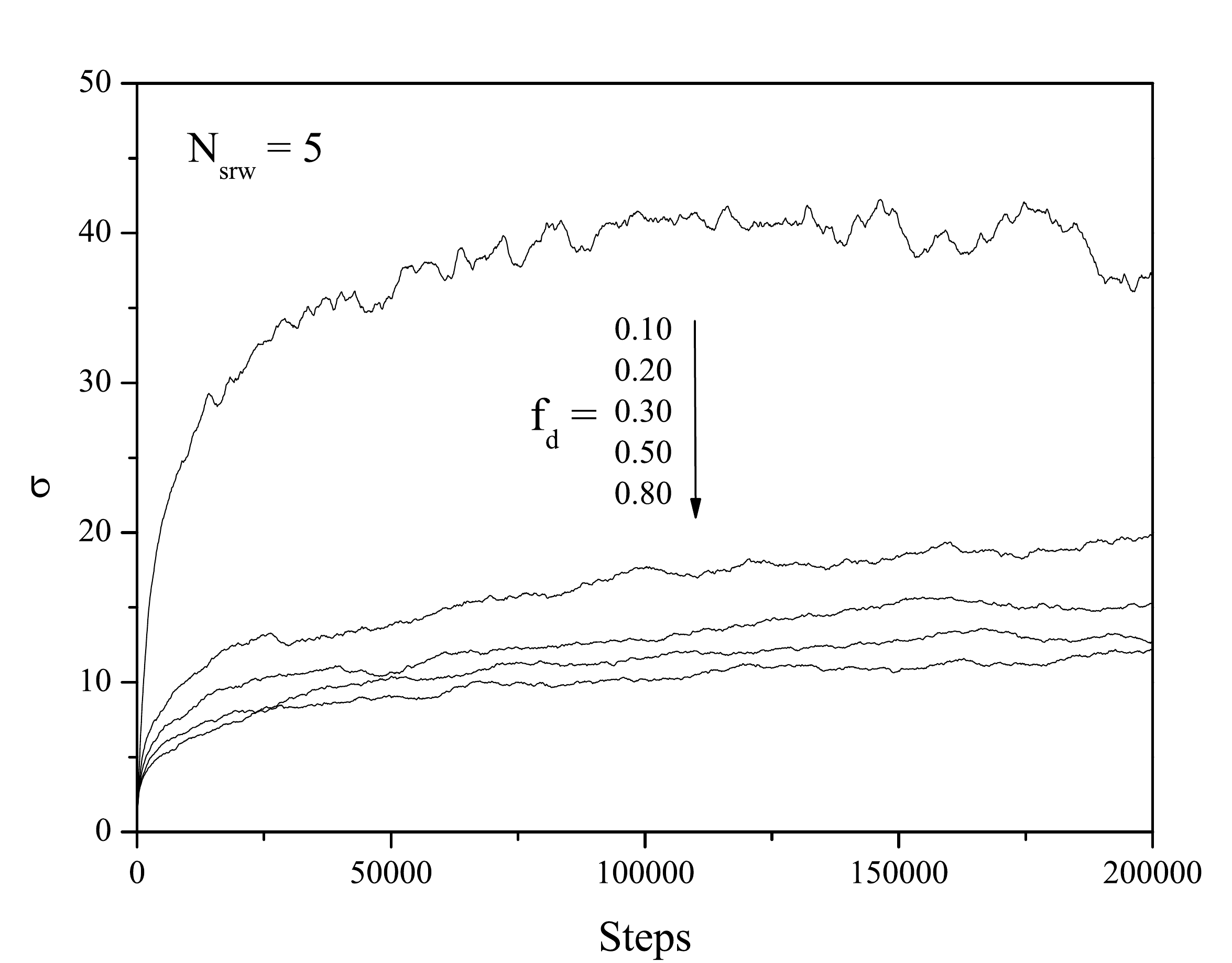} 
\caption{\label{fig:frontwidth} %
The evolution of the front width as given by the root of mean-square deviation from the mean position. The parameters are the
same as in figure~\ref{fig:roughness}.}
\end{center}
\end{figure}

A constant value for the roughness factor $r$ means that the length of the FL is constant but nevertheless different shapes can be observed for the FL. The deformations at a constant length can be visualized with a flexible cord rearranged without stretching or squeezing.
Therefore, the roughness factor $r$ does not reflect all the geometrical features of the FL. Another parameter associated with the roughness is the width of the FL defined as the root of mean-square
deviation $\sigma$ of the FL from its mean position
\cite{barabasi}:
\begin{equation}
\sigma (N_{\text{steps}} ) = \left\{ \frac{1}{N_x }\sum\limits_{i =
1}^{N_x } [h_i - h(N_{\text{steps}} )]^2 \right\}^{1 / 2},
\label{sigma}
\end{equation}
where $h_i$ is the height of the pile of material sites in the
$i$-th column from the bottom of the simulation box up to the
first environment site in this column. The mean position of the
FL is defined as usual \cite{barabasi}:
\begin{equation}
h(N_{\text{steps}} ) = \frac{1}{N_x }\sum\limits_{i = 1}^{N_x } {h_i }\,. %
\label{shift}
\end{equation}
The dependence of $\sigma(N_{\text{steps}})$ on the number of simulation
steps is shown in figure~\ref{fig:frontwidth}. As expected, the average width
of the front profile increases when the density of
ICs in the system decreases. Qualitatively, the evolution of $\sigma$ is
similar to that of $r$, but the convergence to the stationary
regime is noticeably slower.

\begin{table}[!b]
\begin{center}
\caption{\label{tab:slope}The slopes of the dependencies of the mean position of FL
for the same parameters as in table~\ref{tab:size}.}
\vspace{2ex}
\begin{tabular}
{|c|c|c|} \hline\hline \raisebox{-1.50ex}[0cm][0cm]{$f_{\text d}$}& $N_{\text{srw}}= 3$ &
$N_{\text{srw}}= 5$\\
\cline{2-3}
 &
$\times10^{-3\strut}$&
$\times10^{-3}$ \\
\hline\hline ~0.10~& --&
19.12 \\
\hline ~0.20~& 13.24&
\textbf{13.74} \\
\hline ~0.30~& \textbf{9.92}&
\textbf{12.51} \\
\hline ~0.50~& \textbf{8.60}&
\textbf{11.70} \\
\hline ~0.80~& \textbf{8.05}&
\textbf{11.32} \\
\hline\hline
\end{tabular}
\end{center}
\end{table}

\begin{figure}[!b]
\vspace{-3mm}
\begin{center}
\includegraphics[clip,width=0.49\textwidth,angle=0]{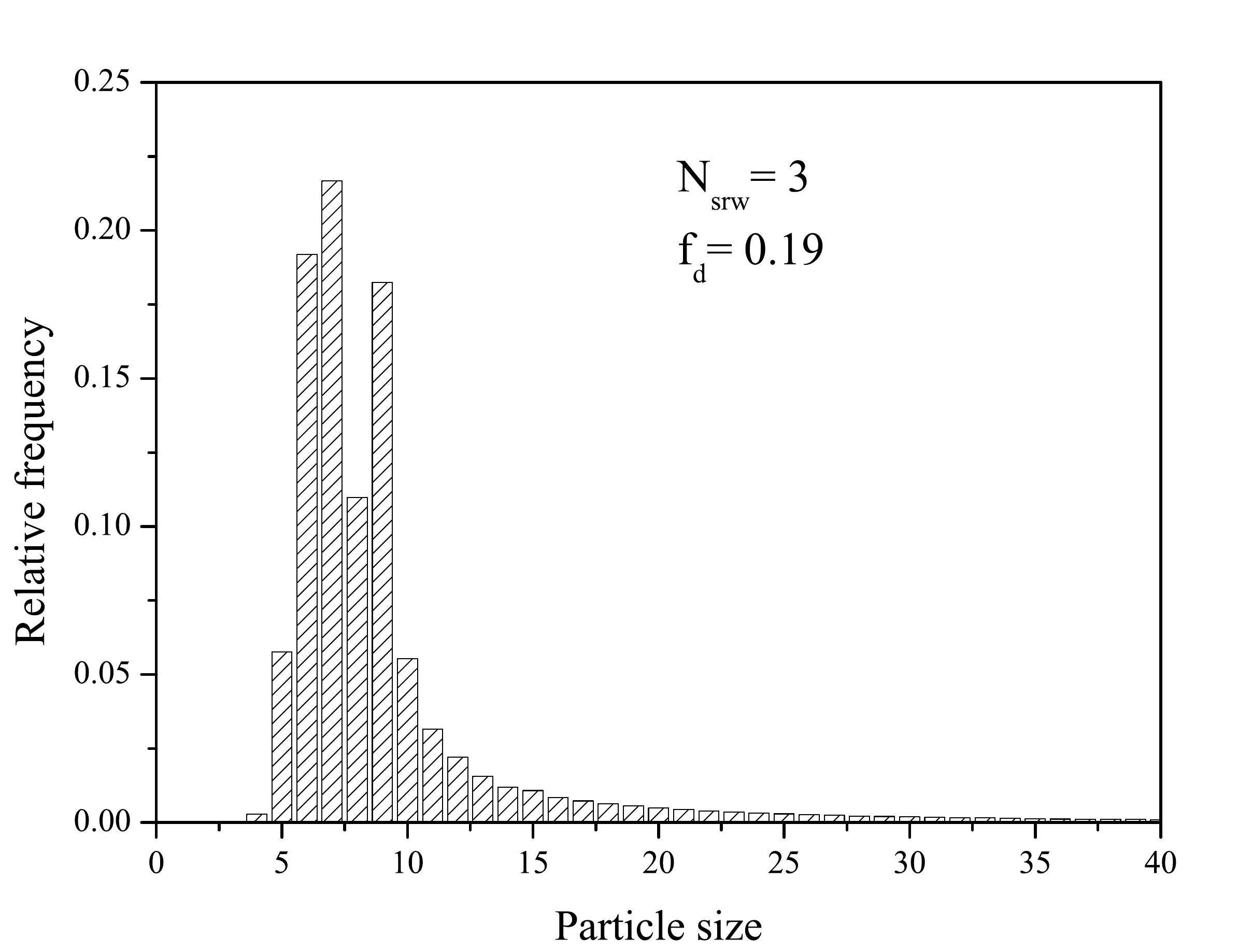}
\includegraphics[clip,width=0.49\textwidth,angle=0]{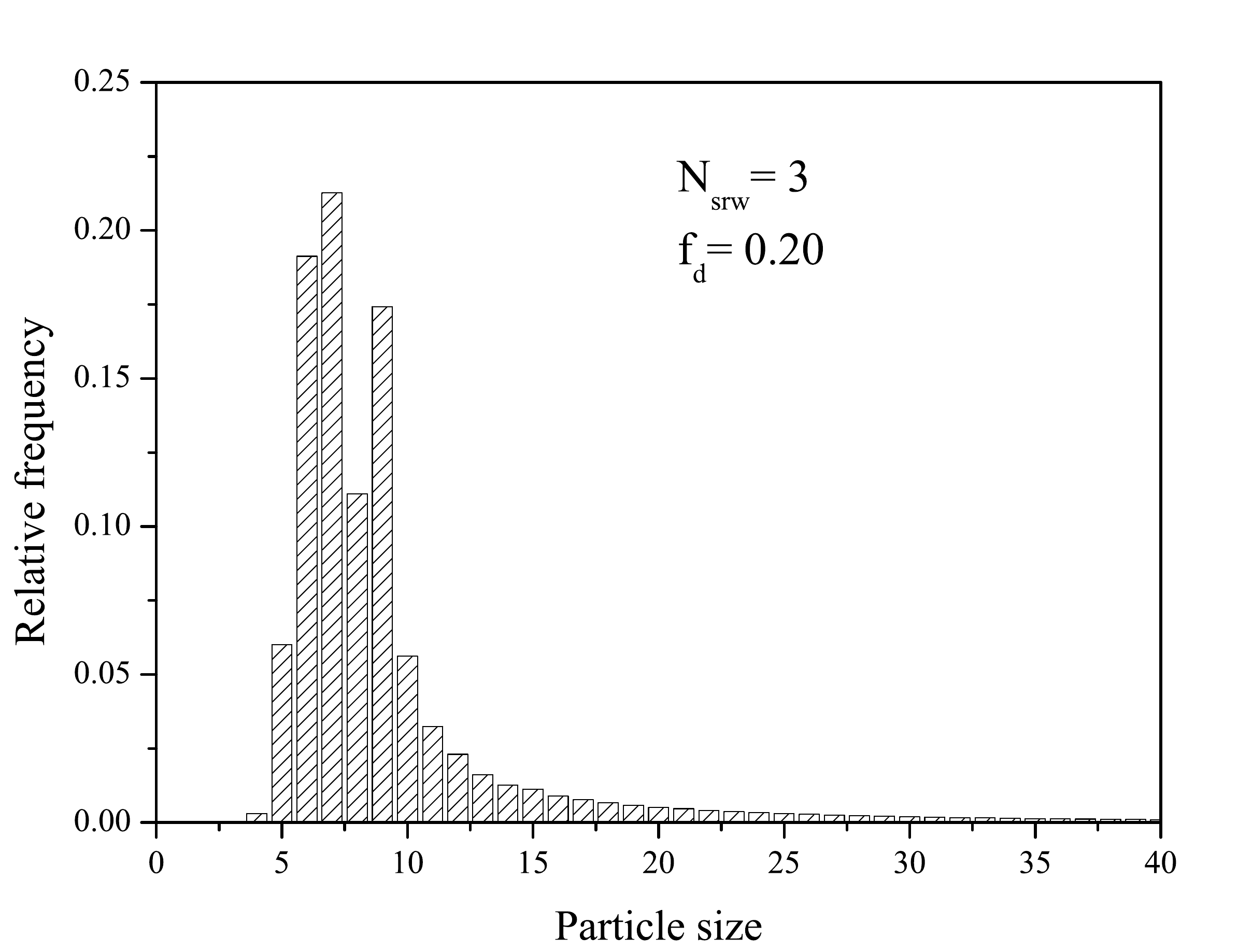}
\includegraphics[clip,width=0.49\textwidth,angle=0]{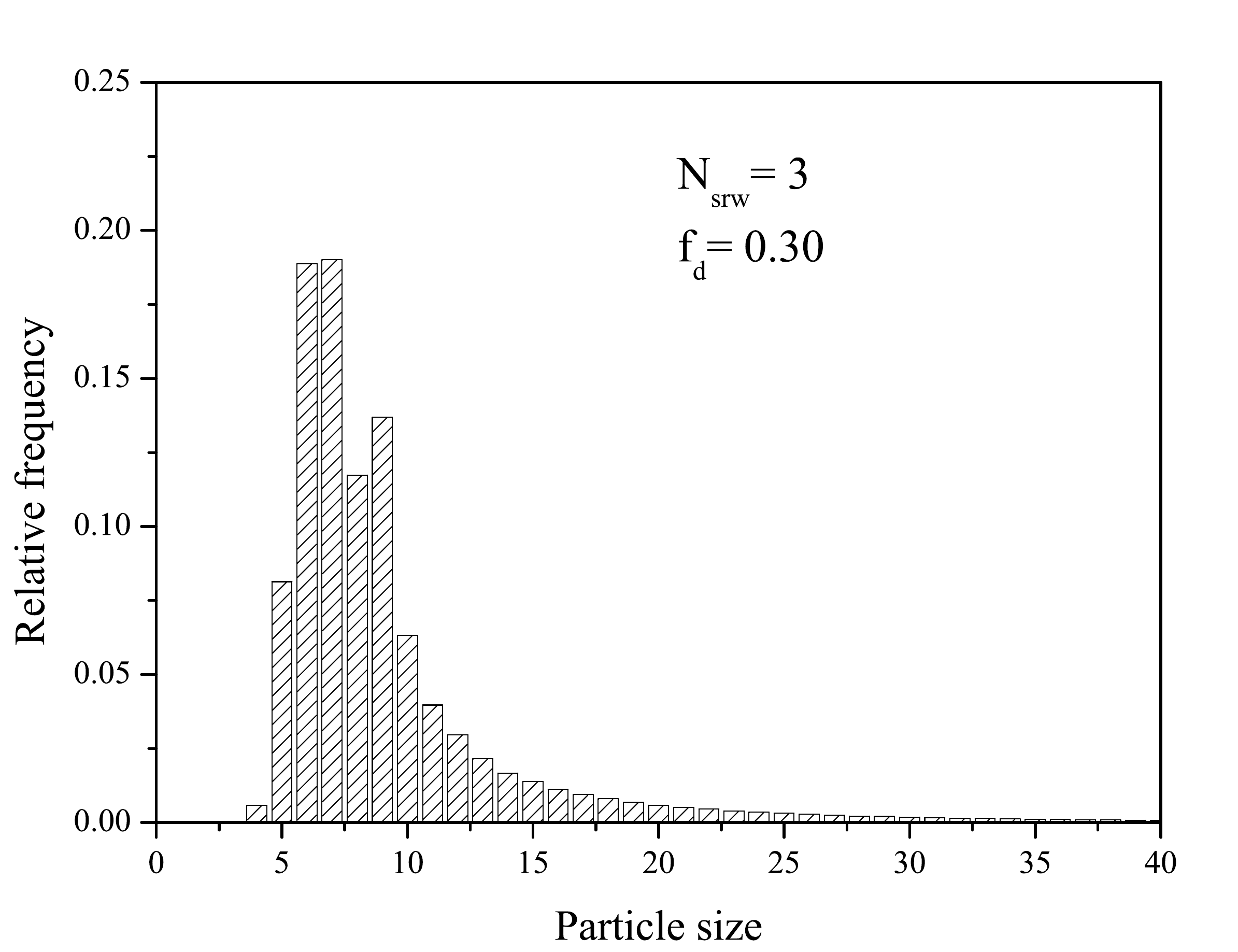}
\includegraphics[clip,width=0.49\textwidth,angle=0]{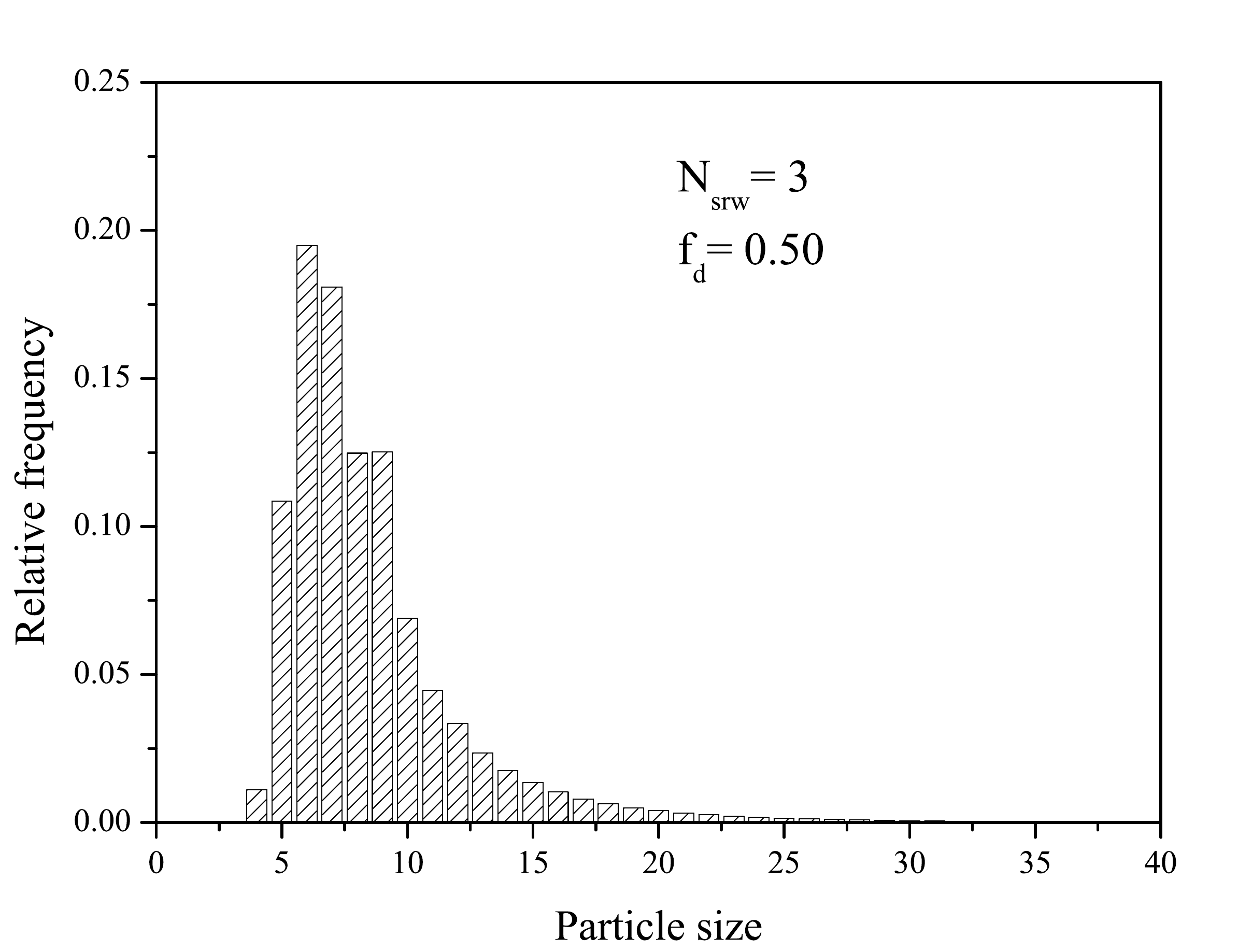} 
\caption{\label{fig:histo3} %
The distributions of the emitted particles for $N_{\text{srw}}=3$ and different densities $f_{\text d}$.}
\end{center}
\end{figure}

The time evolution of the mean position of the FL calculated from
the expression (\ref{shift}) gives an estimation of degradation rate for the material. In our
algorithm, one detachment of the particle occurs per one
simulation step. The simulation step can be related to real time
if we know the so-called dust productivity parameter which is known in some experiments.
Unfortunately, we have no such parameter in the case of the LFCM.
Therefore, we limit the analysis of the FL to its dependence on the number of simulation
steps. We observe the existence of a stationary regime for the displacement of the FL. After a large number of simulation steps, the mean position of FL is given by $p N_{\text{step}}$ and the slopes $p$ have been estimated (table~\ref{tab:slope}). It is observed that the lower is the value of $f_{\text d}$ the larger is the slope.
Due to a single particle detachment per a simulation step, the higher slopes
indicate that the larger sizes of particles are detached. As shown above, the lower is $f_{\text d}$ the
higher is the roughness and the larger are the sizes of the particles emitted. Therefore,
the results presented in table~\ref{tab:slope} are consistent with that in table~\ref{tab:size}.
Indeed, they are in a very good agreement for the values of $f_{\text d}$ higher than $0.30$ and $N_{\text{srw}}=3$ as well as for $f_{\text d}$ higher than $0.20$ and $N_{\text{srw}}=5$ (marked bold). Near the
critical values of ICs density, $f_{\text c}$, there is a difference observed between these two cases.

\begin{figure}[!b]
\vspace{-3mm}
\begin{center}
\includegraphics[clip,width=0.49\textwidth,angle=0]{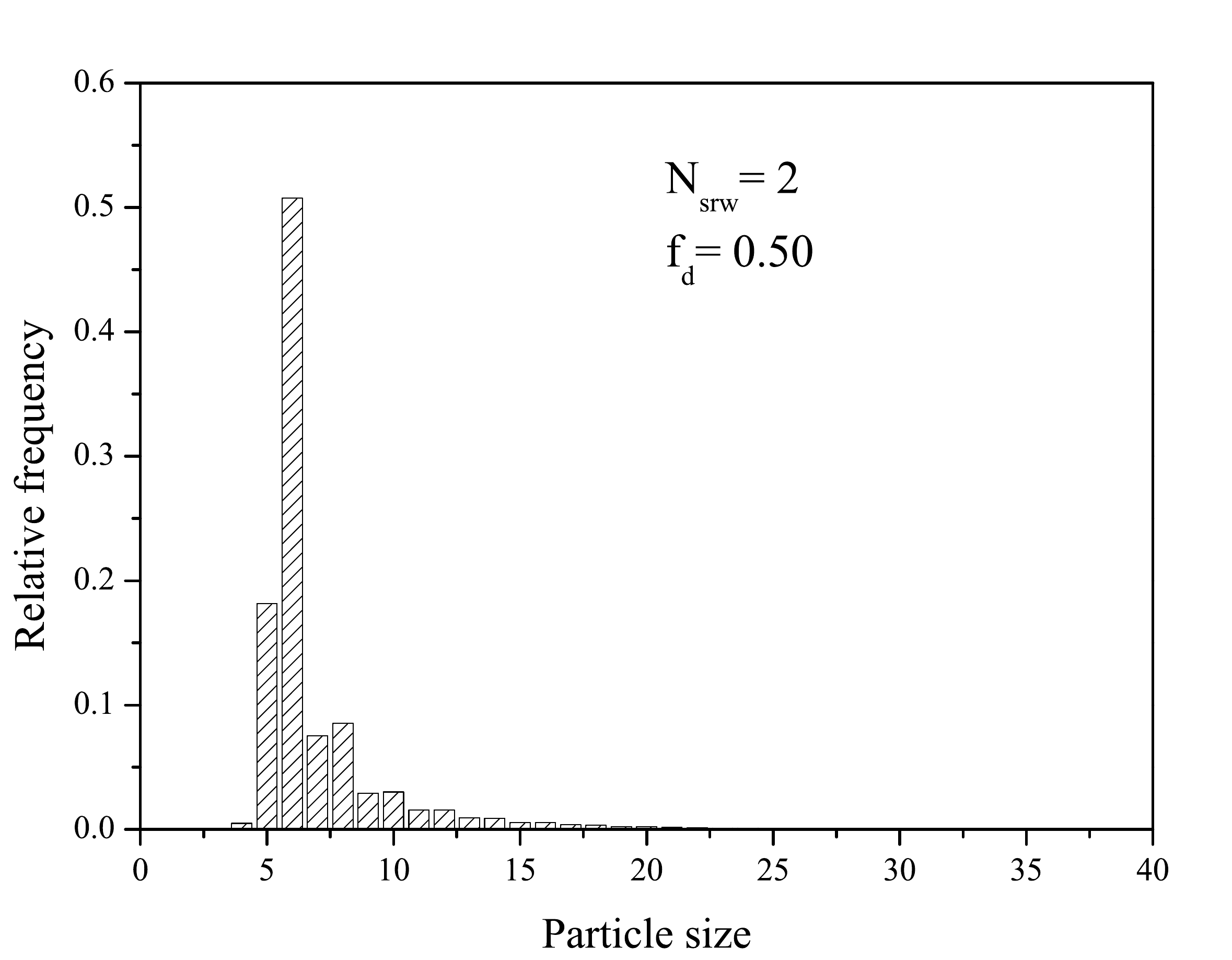}
\includegraphics[clip,width=0.49\textwidth,angle=0]{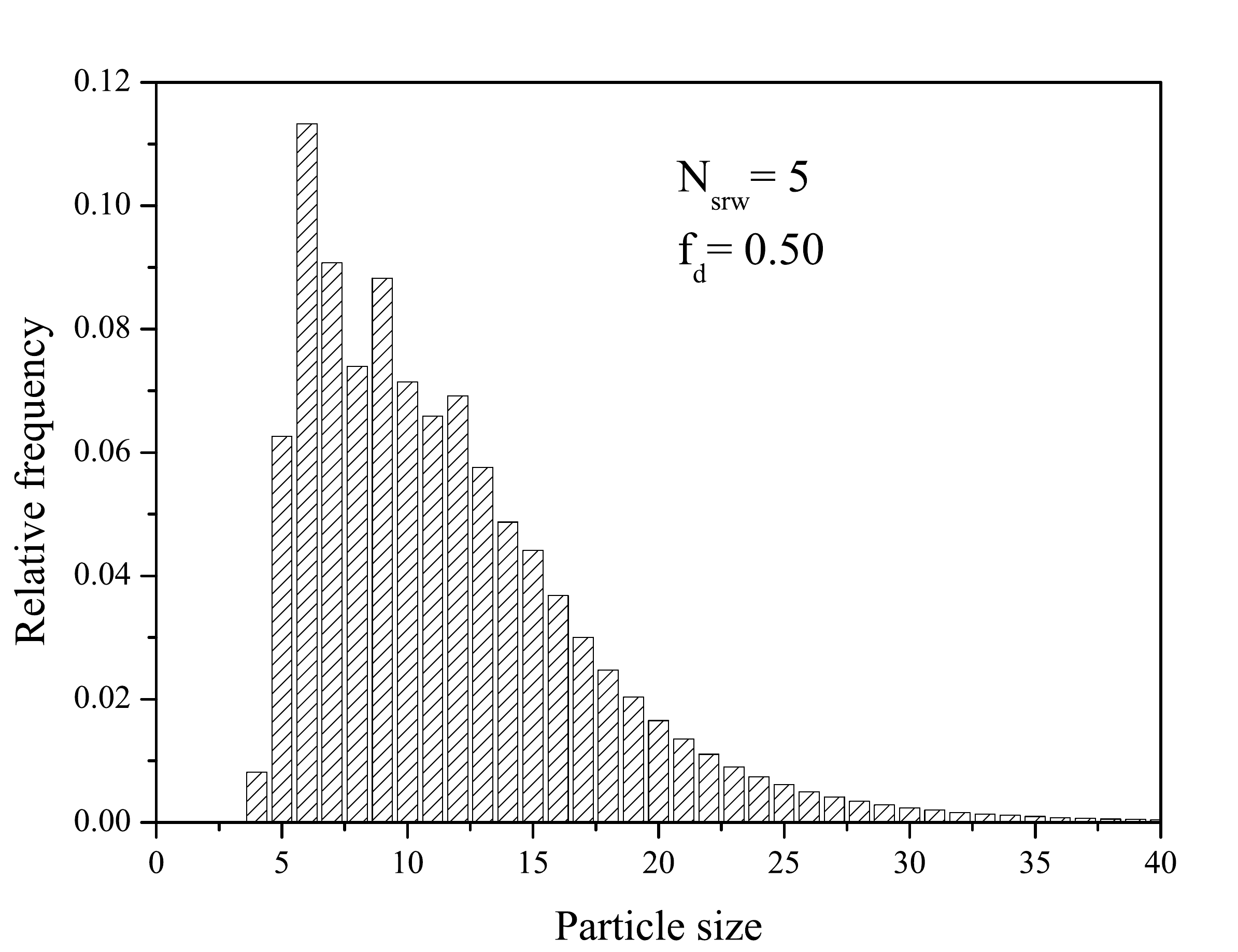} 
\caption{\label{fig:histo25} %
The distributions of the emitted particles for $N_{\text{srw}}=2$ and $N_{\text{srw}}=5$ at the density $f_{\text d}$=0.50.}
\end{center}
\end{figure}

The events for the size histogram are collected in a stationary regime of the material destruction after $50000$ simulation steps. From the simulations we can calculate the total number of the emitted particles and estimate the relative frequency of a given particle size. By such procedure, the histogram is normalized and from this histogram we may determine the mean size of the emitted particles (table~\ref{tab:size}).
In figures~\ref{fig:histo3} and \ref{fig:histo25} we present the calculated histograms. One can see that the dispersion of the particle sizes is not essentially affected by the IC density while
the mean size of the detached particles depends on $f_{\text d}$ as it is shown in
table~\ref{tab:size}. At low $f_{\text d}$, there are two sharp peaks on the
histograms (figure~\ref{fig:histo3}). When $f_{\text d}$ increases, the second
peak is suppressed and the histogram becomes smoother. At a
sufficiently high density ($f_{\text d}=0.50$), the second peak disappears completely.
Simultaneously, the first peak shifts to smaller sizes. The
dispersion of the size distribution is strongly affected by the
$N_{\text{srw}}$ parameter. In figure~\ref{fig:histo25}, the histograms for the
cases of $N_{\text{srw}}=2$ and $N_{\text{srw}}=5$ at $f_{\text d}=0.50$ are presented.
For $N_{\text{srw}}=2$, the dispersion of particle sizes is much smaller
than for $N_{\text{srw}}=5$. Hence, for longer fractures we may obtain larger particles.

\begin{figure}[!b]
\begin{center}
\includegraphics[clip,width=0.49\textwidth,angle=0]{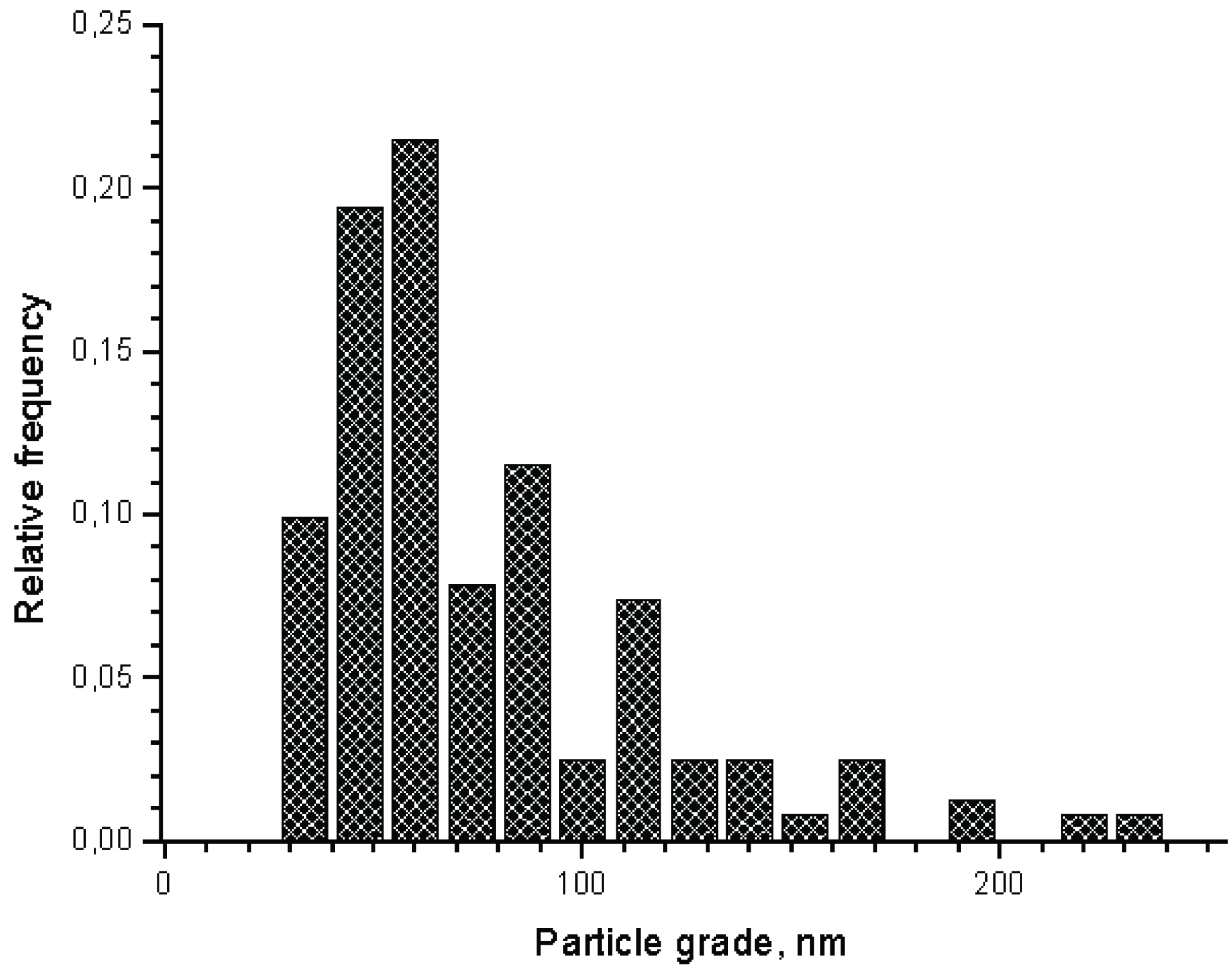}
\caption{\label{fig:histoexp} %
The distribution of the emitted particles obtained from the
experiment for the material containing a lava-like nuclear fuel by~Baryakhtar~et~al.~\cite{baryakhtar}.}
\end{center}
\end{figure}

The projected histogram calculated for $f_{\text d}=0.19{-}0.3$ and $N_{\text{srw}}=3$
(figure~\ref{fig:histo3}) is in a remarkable qualitative agreement with
that shown in figure~\ref{fig:histoexp} for the data of Baryakhtar  et
al. \cite{baryakhtar}. Herein above we have given the relation between the effective radius, the mean number of sites and the lattice constant $a$. If we fit the effective radius to the value $R_{\text{max}}$ associated with the maximum of the experimental histogram and if  for the number of sites we take its value {$n_{\text{max}}$} obtained from the calculated histogram, we can
get an estimation of the lattice constant as
\begin{equation}
a \approx R_{\text{max}}\sqrt{\piup/{n_{\text{max}}}} = {20.1}~\text{nm}\,,
\end{equation}
where $R_{\text{max}}=D/2$, while $D=60$~nm is the diameter of the
particles at the first maximum in figure~\ref{fig:histoexp}, and ${n_{\text{max}}}=7$ is
the corresponding size of the particles in terms of a number of
sites obtained from the simulation. The calculated value of $a$ has the same order of magnitude as the damaged regions $25{-}30$~nm observed in
\cite{baryakhtar} and related to a size of the elementary cluster
that can be detached. The value of the parameter $N_{\text{srw}}$ giving the maximum {length of the fracture}
can be derived as $l=La=(2 N_{\text{srw}}+1)a\sim {141}$~nm for $N_{\text{srw}}=3$.
The value of $l$ is far lower than the critical one when the fracture starts to propagate
continuously~\cite{stashchuk}. Due to this,
the immediate destruction of the material only appears in its subsurface
region or {when fractures form an infinite network inside the
material bulk. In the given study, we consider the first case.}
As it is shown above, the surface roughness keeps
constant during this process. For a more thorough verification of
the model, a comparison of the experimentally determined and
simulated surface roughness can be useful. Such results would
allow us to adjust the model parameters  so far undetermined in the
absence of the experimental data.

\section{Conclusions}
\label{sec6}
We propose a stochastic model to simulate the destruction processes
in nuclear fuel containing (NFC) materials. In this study, we focus on the surface erosion processes
leading to the emission of material particles.
It is assumed that the erosion of the NFC material surface results from numerous $\alpha$-decay events
in the subsurface region, which are accompanied with heavy nuclear recoils producing damages in the material.
The approach used in the study allows us to run simulations at the mesoscopic level and to
describe the surface evolution in large time scales.
A quantitative description of this model {is presented as its projection on a two-dimensional lattice}.

The model is defined by two parameters, one of which is related to the number density of the nuclear fuel in the material,
and another characterizes a maximum size of the defects (fractures) produced by heavy nuclear recoils.
Different pairs of these parameters have been checked and the corresponding characteristics,
such as surface roughness, mean-square width of the front and the yield of material are calculated.
These results were obtained in a stationary regime of the NFC material destruction process.
However, it should be noted that the stationary regime can be achieved only when the number density
$f_{\text d}$ and the maximum length of fractures $2N_{\text{srw}}$ are large enough.
For instance, for each value of the parameter $N_{\text{srw}}$, one can find a critical concentration $f_{\text c}$ below which
the destruction process cannot proceed.
On the contrary, for $f_{\text d}$ higher than $f_{\text c}$, we find at longer times a stationary regime.
It is observed that for high values of $f_{\text d}$, the roughness of NFC material surface is rather small.
However, a decrease of $f_{\text d}$ leads to an increase of the roughness as well as of the mean-square width of the front and
the mean particle size. Moreover, it is noticed that the maximum values of these characteristics
are reached if $f_{\text d}$ tends to $f_{\text c}$.

The size histograms of the particles produced during the erosion of the material surface are built during simulations.
These distributions are characterized by different dispersions and mean values of the particle sizes
depending on $f_{\text d}$ and $N_{\text{srw}}$ chosen. It is shown that our theoretical predictions are in a good agreement
with the experimental histograms of particle sizes obtained due to destruction processes of the material containing lava-like nuclear fuel formed during the Chernobyl catastrophe~\cite{baryakhtar}. We have also checked that the parameters of the model in this case are of a correct order of magnitude.

\section*{Acknowledgement}
The simulations have been done on the computing cluster of the
Institute for Condensed Matter Physics~(Lviv, Ukraine).

\newpage



\ukrainianpart

\title{Стохастичне комп’ютерне моделювання процесів руйнування в самоопромінених матеріалах}
\author{Т. Пацаган\refaddr{icmp}, А. Талеб\refaddr{psl,upmc}, Я. Стафієй\refaddr{cswu}, М. Головко\refaddr{icmp}, \framebox{Ж.П. Бадіалі\refaddr{icmp,upmc}}}
\addresses{
\addr{icmp} Інститут фізики конденсованих систем Національної академії наук України,\\ вул. Свєнціцького, 1, 79011 Львів, Україна
\addr{psl} Дослідницький університет науки та літератури Парижу, Chimie ParisTech --- CNRS, Інститут хімічних досліджень Парижу, Париж, Франція
\addr{upmc} Університет П’єра і Марії Кюрі, Париж, 75231, Франція
\addr{cswu} Університет кардинала Стефана Вишинського, факультет математики і природничих наук,\\ Варшава, Польща
}
\makeukrtitle

\begin{abstract}
\tolerance=3000%
Самоопромінення, що виникає внаслідок процесів поділу радіоактивних елементів, є звичним явищем, яке спостерігається в радіоактивних паливовмісних матеріалах (ПВМ). Численні $\alpha$-розпади призводять до перетворення локальної структури ПВМ. Ушкодження, які виникають за рахунок ударів віддачі осколків поділу важких ядер в приповерхневому шарі, можуть спричинювати відокремлення частинок матеріалу. Така поведінка є подібною до процесу розпилення, яке спостерігається під час бомбардування поверхні матеріалу потоком енергетичних частинок.  Проте, в ПВМ удари ініціюються в об’ємі матеріалу. В цій роботі ми пропонуємо двовимірну мезоскопічну модель, з метою здійснення стохастичного комп’ютерного моделювання процесів руйнування, що виникають в приповерхневій області ПВМ. Ми описуємо ерозію поверхні матеріалу, зміну її шорсткості з часом і передбачаємо відокремлення частинок матеріалу. У цьому дослідженні отримано розподіли розмірів відокремлених частинок. Результати комп’ютерного моделювання якісно узгоджуються із гістограмами розмірів розпилених частинок спостережених в лаво-подібних паливовмісних масах, які сформувалися під час Чорнобильської катастрофи.
\keywords радіоактивні паливовмісні матеріали, самоопромінююче ушкодження, руйнування, шорсткість, стохастичне комп’ютерне моделювання, процес розпилення
\end{abstract}

\end{document}